\documentstyle[aps,prl,amssymb]{revtex}
\title{Conditions for the confirmation of three-particle non-locality }

\author{Peter Mitchell$^{1}$, Sandu Popescu$^{1,2}$ and David Roberts$^{1}$}

\address{ $^1$H.H. Wills Physics Laboratory,
University of Bristol, Tyndall Avenue, Bristol BS8 1TL, UK
\\ $^2$ Hewlett-Packard Laboratories,
 Stoke Gifford, Bristol BS12 6QZ, UK}
\date{19 Nov  2001}
\begin{document}
\maketitle
\begin{abstract}
The notion of genuine three-particle non-locality introduced by
Svetlichny \cite{Svetlichny} is discussed. Svetlichny's inequality
which can distinguish between genuine three-particle non-locality
and two-particle non-locality is analyzed by reinterpreting it as
a frustrated network of correlations. Its quantum mechanical
maximum violation is derived and a situation is presented that
produces the maximum violation. It is shown that the measurements
performed in recent experiments to demonstrate GHZ entanglement
\cite{Bouwmeester}, \cite{Pan} do not allow this inequality to be
violated, and hence can not be taken as confirmation of genuine
three-particle non-locality. Modifications to the experiments that
would make such a confirmation possible are discussed.

\end{abstract}

\pacs{PACS numbers:}

\newcommand{\tr}{\mbox{Tr} }
\newcommand{\ket}[1]{\left | #1 \right \rangle}
\newcommand{\bra}[1]{\left \langle #1 \right |}
\newcommand{\amp}[2]{\left \langle #1 \left | #2 \right. \right \rangle}
\newcommand{\proj}[1]{\ket{#1} \! \bra{#1}}
\newcommand{\ave}[1]{\left \langle #1 \right \rangle}
\newcommand{\superop}{{\cal E}}
\newcommand{\unity}{\mbox{\bf I}}
\newcommand{\hilbert}{{\cal H}}
\newcommand{\relent}[2]{S \left ( #1 || #2 \right )}
\newcommand{\banner}[1]{\bigskip \noindent {\bf #1} \medskip}
%%%  Mathematical abbreviations  %%%
\newcommand{\I}{{\mathbf I}}
\newcommand{\R}{{\mathbf R}}
\renewcommand{\S}{{\mathbf S}}
\newcommand{\up}{\uparrow}
\newcommand{\down}{\downarrow}

\section{Introduction}

Generalized Bell inequalities have been reported for N-particle
systems which show that quantum mechanics violates local realism
in these situations \cite{Popescu} \cite{Gisin} \cite{Mermin}.
However such results are insufficient to show that all of the
particles in a system are acting non-locally -  it is possible to
imagine a non-local many-particle system as consisting of a finite
number of non-local sub-systems, but with only local correlations
present between these  sub-systems. For example a state of three
particles $\ket{\Psi}_{123}$ which can be decomposed as
$\ket{\psi}_{1}\ket{\phi}_{23}$ only exhibits non-local
correlations between particles 2 and 3. Similarly, a density
matrix $\rho_{123}$ which is a mixture of states of the form
$\ket{\psi}_{1}\ket{\phi}_{23}$, $\ket{\eta}_{2}\ket{\xi}_{13}$
and $\ket{\chi}_{3}\ket{\theta}_{12}$ contains only two particle
non-locality (though it might be very difficult to show this if
only the density matrix is given but not the explicit
decomposition). Suppose however that $\ket{\Psi}_{123}$ can not be
decomposed - does this necessarily imply that it has three
particle non-locality?  This question was first raised by G.
Svetlichny \cite{Svetlichny}. More precisely, Svetlichny asked the
following: We know that the correlations between the results of
measurements performed on triplets of particles in the state
$\ket{\Psi}_{123}$ cannot be described by local hidden variables.
Could they however be described by a {\em hybrid local - nonlocal}
system, in which non-local correlations are present only between
two particles (which two particles are nonlocally correlated can
change in different runs of the experiment) while they are only
locally correlated with the third ? If ``yes" than although
$\ket{\Psi}_{123}$ can not be decomposed as a direct product of
one particle versus a (possible entangled) state of the other two,
the nonlocality exhibited by this state is still only two particle
nonlocality.

Formally Svetlichny's model is the following. Let $P(A=a, B=b,
C=c)$ be the probability for obtaining a results $A=a$, $B=b$ and
$C=c$ when observable $A$ is measured on the first particle, $B$
on the second and $C$ on the third. In a local hidden variables
model each particle in the triplet is endowed at source with the
same hidden variable $\lambda$ and later, when subjected to
measurements, each particle behaves independently of the others,
taking into account only the value of the hidden variable and the
measurement to which itself is subjected, but not to what
measurements the other particles were subjected and/or the results
they yield. Hence, $P(A=a, B=b, C=c)$ can be expressed as

\begin{equation}
P(A=a, B=b, C=c)_{local}= \int \rho(\lambda)d\lambda
P_1(A=a|\lambda) P_2(B=b|\lambda) P_3(C=c|\lambda),
\end{equation}
where $\rho(\lambda)$ describes the probability that the hidden
variable has a particular value $\lambda$. As it is well-known no
such local hidden variables model can account for the correlations
generated by entangled states.

In the hybrid local-nonlocal hidden variables model considered by
Svetlichny,  $P(A=a, B=b, C=c)_{Sv}$ is given by:

\begin{eqnarray}
P(A=a, B=b, C=c)_{Sv}=q_{12} \int \rho_{12}(\lambda)d\lambda
P_{1,2}(A=a,B=b|\lambda) P_3(C=c|\lambda)\nonumber\\ +q_{23} \int
\rho_{23}(\lambda)d\lambda
P_{2,3}(B=b,C=c|\lambda)P_1(A=a|\lambda)\\ +q_{13} \int
\rho_{13}(\lambda)d\lambda
P_{1,3}(A=a,C=c|\lambda)P_2(B=b|\lambda), \nonumber
\end{eqnarray}
subject to $q_{12}+q_{23}+q_{13}=1$ and $\int
\rho_{i,j}(\lambda)d\lambda=1$.

Thus when repeated measurements are performed on an ensemble, the
three terms in equation 2 correspond to the three possible
factorizations of two particle non-locality between the three
particles, (1,2)-3, (2,3)-1 and (1,3)-2, with $q_{12}, q_{23}$ and
$q_{13}$ the probabilities of each particular factorization being
present.

Svetlichny derived an inequality which is obeyed by all such
hybrid local - two-particle nonlocal models, and showed that some
quantum states violate the inequality, hence they are genuinely
three-particle nonlocal.

In this paper Svetlichny's inequality is first given a novel
interpretation as a frustrated network of correlations. It is
hoped that such an interpretation will give physical insight into
the subsequent discussions. We then derive the maximal possible
violation of Svetlichny's inequality and a quantum state is then
presented which violates it maximally. Finally we discuss the
experimental status of the verification of genuine 3-particle
non-locality, and suggest simple modifications to the recent
experiments by D. Bouwmeester et al. \cite{Bouwmeester} and Pan et
al. \cite{Pan} which may make such a verification possible.

\section{interpreting generalized Bell inequalities as frustrated networks}

Bell-type inequalities are generally expressed in terms of the
expectation values of observables.  In this section it is shown
how it is possible to interpret generalized Bell inequalities as
frustrated networks of correlations, and accordingly we present an
alternative derivation of Svetlichny's inequality based upon such
an interpretation. (In fact all presently known Bell type
inequalities can be described in such a way \cite{popescu2}, and
this leads to a better understanding of their physical meaning.)

Consider a situation of three spatially separated two dimensional
systems.  System A is subject to one of the measurements $A$ or
$A^{\prime}$, and similarly for systems B and C.  The result of
any measurement is labeled $\pm1$. Suppose $A$, $B$ and $C$ have
been measured. Since the outcomes $a$, $b$ and $c$ can only be
equal to $\pm1$, we have only two possibilities: either $a=bc$ or
$a=-bc$; we refer to the two cases as $A$ being correlated to $BC$
or anti-correlated to $BC$. Furthermore, when $a=bc$ it is also
the case that $b=ac$ and $c=ab$  thus we can talk about
correlation without mentioning explicitly between which
partitions;  similarly for anti-correlation. Define the
probability of correlation, $P_{c}(ABC)$ as the probability that
$A$, $B$ and $C$ are correlated, and $P_{a}(ABC)$ as the
probability that they are anti-correlated. Now consider the
expression

\begin{equation}
\ S=P_{a}(ABC)+P_{a}(ABC^{\prime})+P_{a}(A^{\prime}BC)+
P_{c}(A^{\prime}BC^{\prime})
\-+P_{a}(AB^{\prime}C)+P_{c}(AB^{\prime}C^{\prime})+P_{c}(A^{\prime}B^{\prime}C)+P_{c}(A^{\prime}B^{\prime}C^{\prime})
\label{newineq}\end{equation}

Suppose initially that limited non-locality takes the form that
particles A and B form a non-local subsytem AB and that this
subsystem is locally correlated with particle C.  The other
possible factorizations of the system, A-BC and B-AC, will be
considered later. Then equation (3) corresponds to the network
shown below:

\setlength{ \unitlength}{1cm}

\begin{picture}(16,7)

\put(6.5,6.5){Figure 1}

\put(1,3){$C^{\prime}$} \put(1,4){$C$}
\put(6.5,2){$A^{\prime}B^{\prime}$} \put(6.5,3){$A^{\prime}B$}
\put(6.5,4){$AB^{\prime}$} \put(6.5,5){$AB$}

\multiput(2,3)(0.2,0.1){21}{\circle*{0.1}}
\multiput(2,4)(0.2,0.05){21}{\circle*{0.1}}
\multiput(2,4)(0.2,0){21}{\circle*{0.1}}
\multiput(2,4)(0.2,-0.05){21}{\circle*{0.1}} \thicklines
\put(2,4){\line(2,-1){4}} \put(2,3){\line(1,0){4}}
\put(2,3){\line(4,-1){4}} \put(2,3){\line(4,1){4}}

\thicklines \put(8,3){\line(1,0){3}}
\multiput(8,4)(0.2,0){15}{\circle*{0.1}}

\put(11.5,4){Anti-correlation} \put(11.5,3){Correlation}

\end{picture}

Recall that in our interpretation of Svetlichny's inequality
non-locality between A and B means these particles are regarded as
a composite system.  Hence the outcomes for the paired
measurements $AB, AB^{\prime}, A^{\prime}B$ and
$A^{\prime}B^{\prime}$ are completely unconstrained from each
other.  Furthermore, locality of C versus AB means that for any
local hidden variable model the choice of which of the paired
measurements $AB, AB^{\prime}, A^{\prime}B$ and
$A^{\prime}B^{\prime}$ to make is independent of whether $C$ or
$C^{\prime}$ is measured.

In the most general hidden variable model that can be considered,
for each value of the hidden variable $\lambda$, the measurements
can yield different outcomes, according to the associated
probabilities, such as $P(A=a|\lambda)$.  However, it can be
easily shown that any such model can be re-cast into a
deterministic model in which for each value of $\lambda$ the
outcomes are completely determined, i.e. the probabilities of
obtaining each of the possible measurements is either 0 or 1. In
particular, for each value of $\lambda$ we have a given,
well-defined assignment of $\pm1$ values for $ab$, $ab'$, $a'b$,
$a'b'$, $c$ and $c'$, and the probabilities of correlation and
anticorrelation are either 0 or 1.

Referring to equation (3), one can easily check that for no
assignment of $\pm1$ values for the results of measurements can
all the eight probabilities be equal to 1, nor can all of them be
equal to 0. In fact at least two of the bonds in figure 1 must be
satisfied by any combination of $\pm1$ at the vertices, and only a
maximum of six out of the total of eight bonds may ever be
satisfied. Hence the network is frustrated (in other words not all
links can be simultaneously satisfied) and for every value of
$\lambda$, $2\le S\le 6$. Furthermore, since the the inequality
holds for every value of $\lambda$, it also holds for the average.

As a last step, due to the symmetry under permutation of
particles, the same inequality holds for all 2 versus 1
partitions, and thus for the grand average over all possible
partitions and all assignments of the hidden variable. While the
values of $S$ for given different partitions or particular values
of the hidden variable are not accessible experimentally - indeed,
when performing the measurements on an ensemble of triplets of
particles we don't know what is the hidden variable or the
partition of a particular triplet, the grand average is
experimentally observable. This is Svetlichny's inequality, in a
slightly different form than originally proposed.

The original inequality reads
\begin{equation}
Sv=\left|E(ABC)+E(ABC^{\prime})+E(A^{\prime}BC)-E(A^{\prime}BC^{\prime})+E(AB^{\prime}C)-E(AB^{\prime}C^{\prime})-E(A^{\prime}B^{\prime}C)-E(A^{\prime}B^{\prime}C^{\prime})
\right| \le 4,\label{originalineq}
\end{equation}
where $E(ABC)$ represents the expectation value of the product
$ABC$. This form of the inequality can be easily deduced from
eq.(\ref{newineq}) and the above established bounds for S, by
noting that the expectation values are related to the
correlation/anti-correlation probabilities by
\begin{equation}
E=2P_c-1=1-2P_a
\end{equation}

\section{The predictions of quantum mechanics for three-body systems}

We now derive the maximum possible quantum mechanical violation of
Svetlichny's inequality and show a particular case in which the
inequality is maximally violated.

It is possible to show that $Sv = 4 \sqrt{2}$ is the maximum
possible quantum mechanical violation of Svetlichny's inequality;
this is the equivalent of Cire'son's bound for the CHSH
inequality. For a state $\ket {\psi}$, $Sv$ can be written as:

\begin{equation}
Sv=\left|\bra{\psi}AB(C+C^{\prime})
\ket{\psi}+\bra{\psi}AB^{\prime}(C-C^{\prime})
\ket{\psi}+\bra{\psi}A^{\prime}B(C-C^{\prime})
\ket{\psi}+\bra{\psi}A^{\prime}B^{\prime}(-C-C^{\prime})
\ket{\psi}\right|,
\end{equation}
by replacing in (\ref{originalineq}) the expectation values by
their quantum expression and grouping the terms. Using Schwarz'
inequality we can bound the magnitude of each term:
\begin{equation}
 |\bra{\psi}AB(C+C^{\prime}) \ket{\psi}| \le \sqrt{|{\bra{\psi}ABAB}) \ket{\psi}||\bra{\psi}(C+C^{\prime})(C+C^{\prime}) \ket{\psi}|}
\end{equation}
\begin{equation}
\le
\sqrt{2+\bra{\psi}CC^{\prime}+C^{\prime}C \ket{\psi}},
\end{equation}
where the last inequality obtains since $\bra{\psi}ABAB
\ket{\psi}=\bra{\psi}CC \ket{\psi}=\bra{\psi}C^{\prime}C^{\prime}
\ket{\psi}=1$.  Similar results are found for the other three
terms. If we now let $x=\bra{\psi}CC^{\prime}+C^{\prime}C
\ket{\psi}$, then

\begin{equation}
|Sv| \le 2(\sqrt{2+x})+2(\sqrt{2-x})
\end{equation}
Thus $|Sv| \le 4 \sqrt{2}$ with the maximum absolute value being
attained at $x=0$.

For a GHZ state of three spin 1/2 particles $ \ket { \psi}=
\frac{1}{\sqrt {2}} \left( \ket {\uparrow \uparrow
\downarrow}-\ket {\downarrow \downarrow \uparrow} \right)$, where
$\uparrow$ and $\downarrow$ represent spins polarized ``up" or
``down" along the $z$ axis,  Svetlichny's inequality is violated
if, for example, measurements are made in the xy plane along some
appropriate directions. In this case $E(ABC)=\bra {\psi}
\vec{a}.\vec{\sigma} \otimes \vec{b}.\vec{\sigma}\otimes
\vec{c}.\vec{\sigma} \ket { \psi} = -cos( \alpha + \beta -
\gamma)$, where we labeled the angles from the $x$ axis. The
inequality will be violated  by choosing $ \alpha =
0,\alpha^{\prime} = \frac {- \pi}{2}, \beta =
\frac{\pi}{4},\beta^{\prime} = \frac {-\pi}{4}, \gamma =
0,\gamma^{\prime} = \frac {\pi}{2}$.  Then $Sv = 4 \sqrt{2}$.

\section{Experiments}

Experiments to produce and analyze 3-particle entangled states are
far more difficult than those on 2-particle entangled states which
are now routinely performed. In fact the very first such
experiments have only very recently been performed. Unfortunately
although the beautiful work of Svetlichny  is now more than a
decade old, the notion of genuine 3-particle nonlocality which it
introduced has not been widely known and the experiments on
3-particle entanglement have not been specifically designed to
verify the existence of such correlations. In this section we
revisit the experiments of Bouwmeester et al. \cite{Bouwmeester}
and Pan et al. \cite{Pan}, two of the first experiments to test
3-particle entanglement. We show that the particular measurements
performed in these experiments are such that they do not produce
(according to quantum mechanics) any violations of Svetlichny's
inequality, so that they cannot be used for the verification of
the existence of genuine three-particle non-locality, (although
they prove 3-party entanglement)\footnote{More precisely, the
measurements in \cite{Bouwmeester} and \cite{Pan} do not allow
testing of genuine 3-particle nonlocality via Svetlichny's
original inequality.  We don't know whether these particular
measurements could violate some other similar inequality, or they
can be modeled by a limited 2-particle nonlocal model, and hence
they are useless.}. Ironically, history repeats itself. The pre
1964 measurements performed in order to establish the existence of
entanglement, though able to confirm entanglement, turned out to
be precisely those that were not appropriate for testing Bell's
inequalities!

The two experiments described in \cite{Bouwmeester} and \cite{Pan}
use essentially the same experimental set-up to produce the
three-photon entangled state $\ket { \psi}= \frac{1}{\sqrt {2}}
\left( \ket {HHV}-\ket {VVH} \right)$. Here H represents
horizontal polarization and V vertical polarization. To verify
that indeed such a GHZ state had been produced, different tests
were made.

It is simpler to represent the state in the z basis writing $\ket
{H}=\ket{\uparrow}$ and $\ket {V}=\ket{\downarrow}$.  Then  $ \ket
{ \psi}= \frac{1}{\sqrt {2}} \left( \ket {\uparrow \uparrow
\downarrow}-\ket {\downarrow \downarrow \uparrow} \right)$. In
\cite{Bouwmeester} measurements (of the optical equivalent) of
spin in the z and x directions were performed. In the subsequent
experiment \cite{Pan} measurements along z, x and y were
performed.

Unfortunately, as it is straightforward to check, measurements
along x,y and z do not lead to Svetlichny inequality violations
for the GHZ state, so the analysis of the data already obtained
cannot prove the existence of genuine 3-party correlations. On the
other hand, it is easy to modify the experiments so as to produce
a maximum violation of Svetlichny's inequality. It is sufficient
to make measurements in the xy plane using the angles listed above
in section three. Where to measure an angle $\theta$ in this plane
it is necessary to perform a measurement which has eigenvectors
$\frac{1}{\sqrt{2}} (\ket{\uparrow}+e^{i\theta}\ket{\downarrow})$
and $\frac{1}{\sqrt{2}}
(\ket{\uparrow}-e^{i\theta}\ket{\downarrow})$, that is
$\frac{1}{\sqrt{2}} (\ket{H}+e^{i\theta}\ket{V})$ and
$\frac{1}{\sqrt{2}} (\ket{H}-e^{i\theta}\ket{V})$.  It should then
be possible to confirm that the state produced demonstrates
genuine three-particle non-locality.

\noindent{\bf
Acknowledgments} We would like to thank D. Bouwmeester, D.
Collins, N. Gisin, A. Kent and N. Linden for very useful
discussions.


\begin{thebibliography}{7}
\bibitem{Svetlichny} G. Svetlichny. Phys Rev D 35, 10, 3066
(1987).
\bibitem{Bouwmeester} D. Bouwmeester, J.-W. Pan, M. Daniell, H. Weinfurther, and A. Zeilinger,
Phys. Rev. Lett. 82, 1345 (1999).
\bibitem{Pan} J.-W. Pan, D.Bouwmeester,  M. Daniell, H. Weinfurther, and A.
Zeilinger, Nature 403, 515 (2000).
\bibitem{Popescu} S. Popescu, D. Rohrlich, Phys. Lett. A, 166, 293
(1992).
\bibitem{Gisin} N. Gisin, H. Bechmann-Pasquinucci, Phys.Lett. A246, 1 (1998).
\bibitem{Mermin} N D. Mermin. Phys Rev Lett 65, 1838 (1990).
\bibitem{popescu2} S. Popescu et al., in preparation.
\end{thebibliography}
\end{document}